\documentclass[12pt,a4paper]{article}
\usepackage[utf8x]{inputenc}
\usepackage{ucs}
\usepackage[english]{babel}
\usepackage{amsmath}
\usepackage{amsfonts}
\usepackage{amssymb}
\usepackage{graphicx}
\usepackage{cite}
\usepackage{url}
\usepackage{authblk}
\usepackage{cite}
\usepackage{url}
\usepackage{float}
\usepackage{color}

\graphicspath{{figures/}}

\usepackage[linktoc=all,colorlinks=true,citecolor=blue,linkcolor=blue]{hyperref}

\usepackage[left=2cm,right=2cm,top=2cm,bottom=2cm]{geometry}

\title{Revealing the phase space structure of Hamiltonian systems using the action}

\author{Francisco Gonzalez Montoya}
\author{Makrina Agaoglou}
\author{Matthaios Katsanikas}
\affil{School of Mathematics, University of Bristol, \\ Fry Building, Woodland Road, Bristol, BS8 1UG, United Kingdom.}

\date{}
\begin{document}
\maketitle

\begin{abstract}

In this work, we analyse the properties of the Maupertuis' action as a tool to reveal the phase space structure for Hamiltonian systems. We construct a scalar field with the action's values along the trajectories in the phase space. The different behaviour of the trajectories around important phase space objects like unstable periodic orbits, their stable and unstable manifolds, and KAM islands generate characteristic patterns on the scalar field constructed with the action. Using these different patterns is possible to identify the skeleton of the phase space and understand the dynamics. Also, we present a simple argument based on the conservation of the energy and the behaviour of the trajectories to understand the values of their actions. In order to show how this tool reveals the phase space structures and its effectiveness, we compare the scalar field constructed with the actions with Poincare maps for the same set of initial conditions in the phase space of an open Hamiltonian system with 2 degrees of freedom.

\end{abstract}

\label{sec:Introduction}

The study of phase space structure is a fundamental problem in dynamical systems. It is essential to understand the trajectories' behaviour and properties from a theoretical and practical perspective. The traditional tools to visualise the phase space structure like the projection of the trajectories in a plane or Poincare maps are important tools to understand many properties of ODE systems' phase space with 3 dimensions. However, the study of the phase space structure of multidimensional systems remains an open and challenging problem.

New tools have been developed to study the phase space structure of multidimensional systems like Fast Lyapunov Exponets (FLE) \cite{Lega2016},  Mean  Exponential Growth Factor of Nearby Orbits (MEGNO) \cite{Cicotta2016}, Smaller  Alignment Indices (SALI),  Generalized  Alignment Indices (GALI) \cite{Skokos2016} and Determinant of Scattering Functions \cite{Drotos2014,Gonzalez2020}. The phase space structure indicators are scalar fields constructed with the system's trajectories. The differences in the scalar field's values give us information of the phase spaces objects that intersect the set of trajectories considered.  

A family of phase space structure indicators that it has been developed recently is the Lagrangian descriptors (LDs) \cite{madrid,Mancho2013,lopesino2017}. Examples of systems that have been analysed, using this method, can be found in \cite{Agaoglou2020,agaoglougar, Main2017, Main2018 ,Main2019, Bartsch2016 ,katsanikas2020detection,katsanikas2020c,Montoya2020}. The most intuitive Lagrangian descriptor is based on trajectories' arc length. The differences in the arc length of the trajectories with nearby initial conditions give us information about the phase space around those initial points. In this work, we take advantage of the Maupertuis' action $S_0$ that defines a natural metric in the phase space of a common class of Hamiltonian systems. With the action, it is possible to construct a Lagrangian descriptor to reveal the phase space structure. 

In Section \ref{sec:A Lagrangian descriptor based on the action}, we explain in more detail the principle that underpins the detection of phase space objects in the phase space using the nearby trajectories. In section \ref{sec:quadratic_normal_form}, we study the Lagrangian descriptor analytically and its behaviour when the trajectories are close to the hyperbolic periodic orbit of the quadratic normal form Hamiltonian with 2 degrees of freedom (DoF). We also explain this result using an intuitive argument based on the conservation of the energy and the behaviour or the trajectories around the unstable hyperbolic periodic orbit. In section \ref{sec:Numerical example}, we apply the Lagrangian descriptor based on the classical action to reveal the phase structures in a non-integrable Hamiltonian system with 2 DoF with unbounded phase space. Finally, in section \ref{sec:Conclusions and remarks} we summarise our conclusions and remarks.

\section{Lagrangian descriptor based on the action}
\label{sec:A Lagrangian descriptor based on the action}

The Lagrangian descriptors, like other chaotic indicators, are scalar fields evaluated in a set of initial conditions in the phase space. The scalar field's values are determined by the behaviour of the trajectories that cross the set of initial conditions. To visualise the principle behind the detection of objects in the phase space we consider a 2 DoF Hamiltonian system with an unstable hyperbolic periodic orbit $\Gamma$. There are two remarkable invariant surfaces that intersect at this hyperbolic periodic orbit \cite{ott,abraham}. In this case, the invariant property means that the trajectories that start on an invariant surface are always contained in the same surface. These two surfaces are called stable and unstable manifold of the unstable hyperbolic periodic orbit $\Gamma$. The definition of the stable and unstable manifolds $W^{s/u} (\Gamma)$ is the following,

\begin{equation}
 W^{s/u} (\Gamma) = \lbrace  \mathbf{x}  \vert \mathbf{x}(t)  \rightarrow  \Gamma, t \rightarrow \pm \infty \rbrace.
\end{equation}

\noindent This means that the stable manifold $W^s(\Gamma)$ is the union of all the trajectories that converge to the periodic orbit $\Gamma$ as the time $t$ goes to $+\infty$. The definition of the unstable manifold $W^u(\Gamma)$ is similar. The unstable manifold is the set of trajectories converging to the periodic orbit as the time $t$ goes to $-\infty$. 

The phase space of a 2 DoF Hamiltonian system has 4 dimensions. For each value of the energy, we can represent the dynamics in a 3 dimensional constant energy manifold. The stable and unstable manifolds $W^{s/u}(\Gamma)$ have two dimensions and form impenetrable barriers that divide the constant energy manifold \cite{Kovacks2001,wiggins3}. Another important property of the stable and unstable manifolds related to the chaotic dynamics is that, if a stable manifold and an unstable manifold intersect transversely at one point, then an infinite number of transversal intersections exist between them. The structure generated by the stable and unstable manifolds is called tangle and defines a set of tubes that direct the phase space dynamics. The trajectories in a tube never cross the boundaries of a tube. This fact is a consequence of the uniqueness of ordinary differential equations' solution and the stable and unstable manifold dimension. 

The trajectories nearby the stable manifold $W^s(\Gamma)$ have similar behaviour just for some time interval. However, those trajectories diverge from the hyperbolic periodic orbit $\Gamma$ after a while. This is a characteristic property of the trajectories in an unstable hyperbolic periodic orbit neighbourhood. Intuitively, the arc length of the trajectories on the stable manifold $W^s(\Gamma)$ grows like the periodic orbit's arc length when the trajectories are close to $\Gamma$. For the other trajectories near $W^s(\Gamma)$, the arc length grows similar only when the trajectories approach $\Gamma$. After the transit through the hyperbolic periodic orbit neighbourhood, the arc length grows differently. This difference makes possible the detection of phase space objects like stable and unstable manifolds of unstable hyperbolic orbits.

Now, we give the general definition of the Lagrangian descriptor. Let us consider a system of ODEs

\begin{equation}
\frac{d \mathbf{x}(t)}{dt} = \mathbf{v}(\mathbf{x}(t)), \quad \mathbf{x} \in \mathbb{R}^n \;,\; t \in \mathbb{R}
\end{equation}

\noindent where the vector field $\mathbf{v}(\mathbf{x}) \in C^r$ ($r \geq 1$) in a neighbourhood of the point $\mathbf{x}$. The values of the Lagrangian descriptor depends on the initial condition $\mathbf{x}_{0} = \mathbf{x}(t_0)$ and on the time interval $[t_0+\tau_{-},t_0+\tau_{+}]$. The Lagrangian descriptor $M$ is defined as,

\begin{eqnarray}
M(\mathbf{x}_{0},t_{0},\tau_{+},\tau_{-}) & = & M_{+}(\mathbf{x}_{0},t_{0},\tau_{+}) + M_{-}(\mathbf{x}_{0},t_{0},\tau_{-}) \nonumber \\
& = & 
\displaystyle{ \int^{t_{0}+\tau_{+}}_{t_{0}} F(\mathbf{x}(t))\; dt + \int^{t_{0}}_{t_{0}+\tau_{-}} F(\mathbf{x}(t)) \; dt ,} \label{eq:LD1}
\end{eqnarray}

\noindent where the function $F$ is any positive function evaluated on the solutions $\mathbf{x}(t)$, $ \mathbf{x}(t_0) = \mathbf{x}_{0}$, and the extremes of the interval of integration $\tau_{+} \geqslant 0$ and $ \tau_{-} \leqslant 0 $ are freely chosen. These times can change between different initial conditions and allow us to stop the integration once a trajectory leaves a defined region in the phase space. In this way, it is possible to reveal only the phase space objects contained in the region considered.

The function $F$ is chosen as positive to accumulate the effects of the trajectories' behaviour. A natural choice is the infinitesimal arc length of the trajectories on the phase space \cite{lopesino2017}. For the detection of phase space objects like stable and unstable manifolds of hyperbolic periodic orbits, in principle, it is possible to use any scalar field generated by the trajectories of the system like the final points of the trajectories in phase space or other quantities related \cite{Gonzalez2012,Drotos2014,Gonzalez2020}. Nevertheless, it is not always trivial to interpret the results systematically.

Notice that the first integral in the Lagrangian descriptor's definition is calculated with trajectories forward in time. Then, it reveals the presence of the phase space objects in the set of initial conditions like stable manifolds. Meanwhile, the second integral is calculated with the backward time and reveals objects like unstable manifolds.

The construction of the Lagrangian descriptor based on the action is as follows. For simplicity, let us consider a Hamiltonian function on Cartesian coordinates

\begin{eqnarray}
H(q_1, \ldots, q_n, p_1, \ldots, p_n) & = & T(p_1, \ldots, p_n) + V(q_1, \ldots, q_n) \;, \quad (q_1, \ldots, q_n, p_1, \ldots, p_n) \in \mathbb{R}^{2n} \nonumber \\ 
                                      & = &  \sum_{i=1}^n \dfrac{p_i^2}{2m_i} + V(q_1, \ldots, q_n) ,\;
\end{eqnarray}

\noindent where $T$ is the kinetic energy and $V$ is the potential energy.
The Maupertuis' action or viva action $S_0$ for a Hamiltonian system is defined as 

\begin{equation}
    S_0 = \int  \sum_{i=1}^n p_i dq_i \quad , \quad p_i\equiv \frac{\partial L}{\partial \dot{q_i}} = m_i \frac{dq_i}{dt} \;, \quad i \in \lbrace 1,\ldots,n \rbrace.
\end{equation}

\noindent where $L$ is the Lagrangian of the system and defines the momentum. Applying the chain rule and the definition of the momentum, we obtain that

\begin{equation}
 \sum_{i=1}^n p_i \; dq_i =  \sum_{i=1}^n m_i \frac{dq_i}{dt} \frac{dq_i}{dt} \; dt    = \sum_{i=1}^n p_i \frac{dq_i}{dt} \; dt = \sum_{i=1}^n  \dfrac{p_i^2}{m_i} \; dt.
 \label{eq:metric}
\end{equation}

Using this identity and Hamilton's equations is possible to write $S_0$ as

\begin{equation}
S_0 = \int  \sum_{i=1}^n p_i dq_i  = \int \sum_{i=1}^n \frac{p_i ^2}{m_i} \; dt = 2 \int T \; dt  = 2 \int ( H - V) \; dt.
\end{equation}

Taking the integral of the kinetic energy $T$ with respect to the time,
the simplest definition for the Lagrangian descriptor based on the action $S_0$ is

\begin{eqnarray}
M_{S_0}(\mathbf{x}_{0},t_{0},\tau_{+},\tau_{-}) 
& = & 
\displaystyle{ 
2 \int^{t_{0}+\tau_{+}}_{t_{0}}  T  \; dt 
+ 2 \int^{t_{0}}_{t_{0}+\tau_{-}} T  \; dt . }\label{eq:LD}
\end{eqnarray}

This result can be generalised to Hamiltonian where the $T$ is a quadratic function of the generalised velocities $(\dot{q_i},\ldots,\dot{q_n})$ and $V$ is only a function of the generalised coordinates $(q_i,\ldots,q_n)$.  The Lagrangian descriptor based on action $M_{S_0}$ is defined by the metric $ds^2= \sum_{i=1}^n m_i dq_i dq_i$, see equation \ref{eq:metric}. More details about this result
and an example where $M_{S_0}$ has been used are in \cite{Montoya2020phase} and \cite{Carpenter2018}. Analytical estimation for a Lagrangian descriptor family based on different variants of ``arc length'' in phase space has been studied in detail in \cite{Wiggins2019finding}. The Lagrangian descriptor based on action $M_{S_0}$ has a similar structure that the elements of this family of Lagrangian descriptors. An elementary exposition about Lagrangian descriptors in Hamiltonian systems is in \cite{LD_book_2020} and more examples are in \cite{LD_aplications_book_2020}.

\section{ Phase space analysis of the  quadratic normal Hamiltonian form using $M_{S_0}$ }
\label{sec:quadratic_normal_form}

An important question for the phase space study is the detection of hyperbolic periodic orbits and their invariant stable and unstable manifolds. This section studies the Lagrangian descriptor's behaviour evaluated on a set of initial conditions that intersect the stable and unstable manifolds of a hyperbolic periodic orbit. The next calculations are similar to the calculations in \cite{Wiggins2019finding}. Let us consider the most simple integrable 2 DoF Hamiltonian system as an initial case for the analysis of the Lagrangian descriptor based on action $M_{S_0}$. The 2 DoF quadratic normal form Hamiltonian is

\begin{equation}
H(x, y, p_x, p_y) = T(p_x,p_y) + V(x,y) = \frac{\omega}{2}(p_x^2 +x^2) +  \frac{\lambda}{2}(p_y^2 - y^2) \;.
\end{equation}

\begin{figure}[htbp]
	\begin{center}
		\includegraphics[scale=0.5]{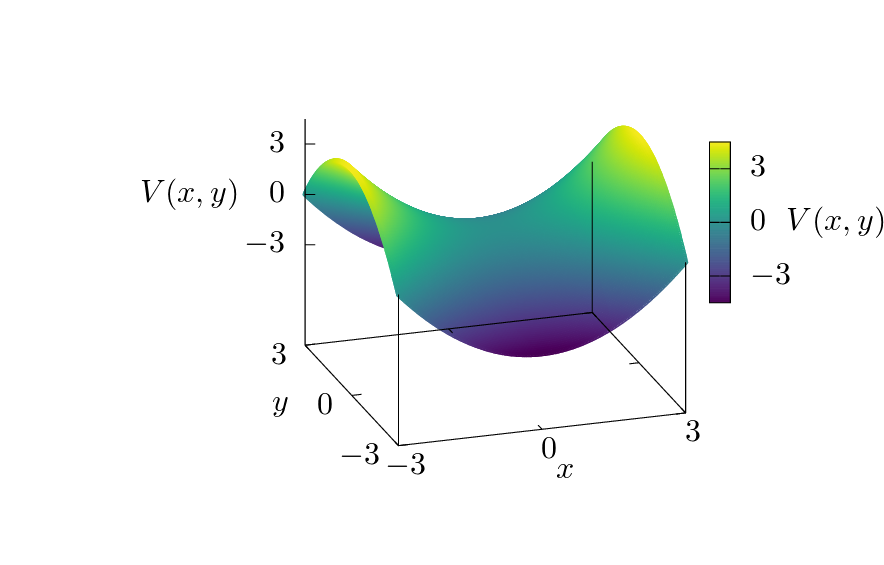}
    \end{center}
    \caption{Potential energy surface $V(x,y)$ corresponding to quadratic normal form Hamiltonian $H(x,y,p_x,p_y)$ with parameters $\omega=1$ and $\lambda =1$. The potential energy surface has index one saddle point at the origin. }
    \label{fig:Vsaddle}
\end{figure}

\begin{figure}[htbp]
    \begin{center}
        \includegraphics[scale=0.5]{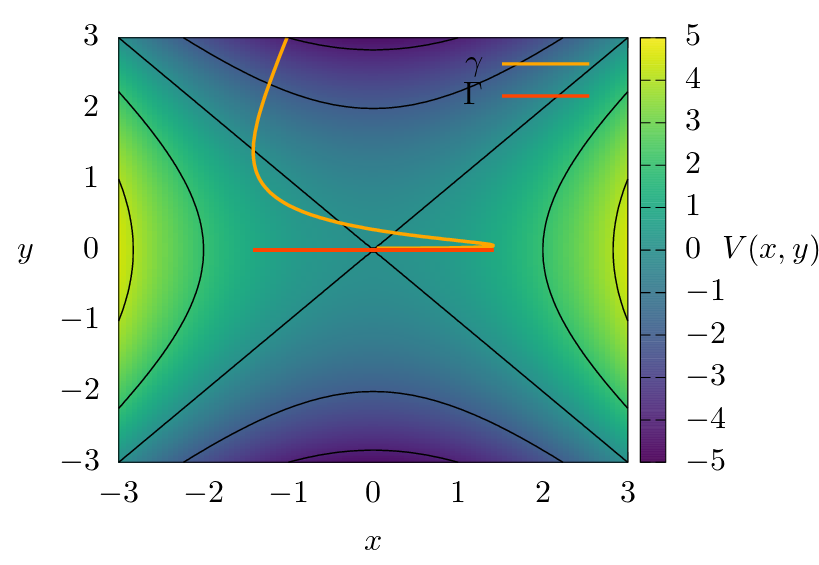}
    \end{center}
    \caption{ Projection of the unstable hyperbolic periodic orbit $\Gamma$ in the configuration space. The potential energy $V(x,y)$ is in colour scale on the background with some equipotential lines on black colour. The red trajectory $\gamma$ is close to the unstable hyperbolic periodic orbit $\Gamma$ for an interval of time before to escape through the region with negative values of $V(x,y)$ to infinity. }
    \label{fig:trajectories_saddle}
\end{figure}

The potential energy surface $V(x,y)$ is saddle with an unstable equilibrium point at the origin, see figures \ref{fig:Vsaddle} and \ref{fig:trajectories_saddle}.

The Hamilton's equations of motion are
\begin{eqnarray}
 \dot{x} &=&\frac{\partial H}{\partial p_x}=\omega p_x,\\
 \nonumber
 \dot{p_x} &=&-\frac {\partial H} {\partial x}=-\omega x,\\
 \nonumber 
 \dot{y}&=& \frac{\partial H}{\partial p_y}=\lambda p_y, \\
 \nonumber
 \dot{p_y} &=&-\frac{\partial H}{\partial y}=\lambda y.\\
 \nonumber
\label{eq}
\end{eqnarray}

The degrees of freedom $x$ and $y$ are uncoupled. The motion in the $x$ component corresponds to a harmonic oscillator and the motion $y$ component to an inverted harmonic oscillator. For this 2 DoF Hamiltonian system exist only one unstable periodic orbit $\Gamma$ that oscillates on the $x$--direction on the line $y=0$ for each value of the energy $E>0$. The orbit $\Gamma$ is a normally hyperbolic invariant manifold and has stable and unstable manifolds. This hyperbolic periodic orbit is given by

\begin{equation}
\label{nhim}
    \Gamma = \{(x,y,p_x,p_y) \;\in \;  \mathbb{R}^4 \;|\; y=p_y=0, E=\frac{\omega}{2} (p_x^2+x^2) \}.
\end{equation}

The unstable and stable invariant manifolds of $\Gamma$ are 
\begin{eqnarray}
\label{man-nhim}
 W^{u}(\Gamma) &=&\{(x,y,p_x,p_y) \;\in \; \mathbb{R}^4 \;|\; y=p_y,E=\frac{\omega}{2}(p_x^2+x^2)\},
 \nonumber\\
 W^{s}(\Gamma)&=&\{(x,y,p_x,p_y) \;\in \; \mathbb{R}^4 \;|\; y=-p_y,E=\frac{\omega}{2}(p_x^2+x^2)\}.
 \nonumber\\
\end{eqnarray}

The Lagrangian descriptor based on the action $M_{S_0}$ for this system is

\begin{eqnarray}
        M_{S_0}(x_0,y_0,p_{x_0},p_{y_0},t_0,\tau_+,\tau_-) &=& M_{S_0}^{e}(x_0,p_{x_0},t_0,\tau_+,\tau_-)  +  M_{S_0}^{h}(y_0,p_{y_0},t_0,\tau_+,\tau_-) \\ \nonumber
    &=& \int_{t_{0}+\tau_+}^{t_{0}+\tau_-}  \frac{\lambda}{2} p^2_x(t,x_0,p_{x_0}) \; dt + \int_{t_{0}+\tau}^{t_{0}+\tau_-} \frac{\omega}{2} p^2_y(t,y_0,p_{y_0}) \; dt  ,
    \label{ld-ini} 
\end{eqnarray}

\noindent where the terms $M_{S_0}^{h}$ and $M_{S_0}^{e}$ are the hyperbolic and the elliptic parts respectively. In the next calculations, we take $t_0=0$ and $\tau_-=0$ to simplify the analysis. 

Starting from the hyperbolic part, the solutions of the equations of motion for $y$ and $p_y$ are

\begin{eqnarray}
\label{sol}
 y(t)= \frac{p_{y_0}}{2}(e^{\lambda t} + e^{-\lambda t}) + \frac{y_{0}}{2}(e^{\lambda t} - e^{-\lambda t} )  ,
 \nonumber\\
 p_y(t)=\frac{p_{y_0}}{2}(e^{\lambda t} - e^{-\lambda t}) + \frac{y_{0}}{2}(e^{\lambda t} + e^{-\lambda t} )  .
 \nonumber\\ 
\end{eqnarray}

In this example, the integral corresponding to the hyperbolic part is

\begin{eqnarray}
    M_{S_{0}}^{h}(y_0,p_{y_0},\tau_+) & = & \int_{0}^{\tau_+}  
    \frac{\lambda}{2} p^2_{y}(t,y_0,p_{y_0}) \; dt
    = \int_{0}^{\tau_+} \frac{\lambda}{2}  (\frac{p_{y_0}}{2}(e^{\lambda t} - e^{-\lambda t}) + \frac{y_{0}}{2}(e^{\lambda t} + e^{-\lambda t}))   \;dt  \nonumber  \\
    &=& (  t \left(- \frac{\lambda^{2} p^{2}_{y_0}}{2} + \frac{\lambda^{2} y^{2}_0}{2}\right) \nonumber \\  &+& \frac{1}{64} \left(- 8 \lambda p^{2}_{y_0} + 16 \lambda p_{y_0} y_0 - 8 \lambda y^{2}_0\right) e^{- 2 \lambda t} \nonumber \\ &+& \frac{1}{64} \left(8 \lambda p^{2}_{y_0} + 16 \lambda p_{y_0} y_0 + 8 \lambda y^{2}_0\right) e^{2 \lambda t} )|^{\tau_+}_{0}.
\end{eqnarray}

This integral grows exponentially with time.  $M_{S_{0}}^{h}$ has a minimum  \cite{LD_book_2020} and in the limit $\tau_+ \rightarrow \infty$, the minimum converge to the initial conditions on the line $y_0=-p_{y_0}$ contained in the stable manifold, see equation \ref{man-nhim}. Analogous result follows for the integration backwards on time and initial condition on the line $y_0=p_{y_0}$.

Now we calculate the elliptic part of the Lagrangian descriptor $M_{S_{0}}^{e} (x_0,p_{x_0},\tau_+)$. The solutions of the equations of motion for harmonic oscillator are

\begin{eqnarray}
x(t)=p_{x_0}\sin(\omega t)+x_0 \cos(\omega t),  
\nonumber\\
p_x(t)=p_{x_0}\cos(\omega t)-x_0 \sin(\omega t). 
\nonumber\\
\end{eqnarray}

Then, we obtain

\begin{equation}
    M_{S_{0}}^{e}(x_{0},p_{x_0},\tau_+) = \int_{0}^{\tau_+} \frac{\omega}{2} {p}^2_{x}(t,x_{0},p_{x_0})\;dt = \int_{0}^{\tau_+} \frac{\omega}{2} (p_{x_0}\cos(\omega t)-x_0 \sin(\omega t))^{2} \;dt.
\end{equation}

If we consider that $p_x$ has period $P=2 \pi/\omega$, then $\tau_+ = NP+r$, where $N$ is an integer and $r\in [0,P]$.  We calculate the integral starting on the initial condition $p_{x_0}=0$ and $x_0=\sqrt{2E/\omega}$ without loss of generality. This gives us  

\begin{eqnarray}
     M_{S_{0}}^{e}(x_{0},p_{x_0},\tau) = E N \int_{0}^{2\pi} \sin^{2}u \;du +\int_{0}^{r}  \frac{\omega}{2} p_x^{2} \;dt=\\
     \nonumber
     2\pi E N  + \int_{0}^{r} \frac{\omega}{2} p^2_{x} \;dt = 2\pi E N +M^{e}_{S_0}(x_0,p_{x_0},r).
     \nonumber
\end{eqnarray}

From the above equation, we see that the elliptic part $M^e_{S_0}$ in every oscillation accumulates the same value of action, in contrast to the hyperbolic component $M^h_{S_0}$ that grows exponentially.  

Considering the results for the hyperbolic and elliptic components of the Lagrangian descriptor based on the action, we can conclude that $M_{S_0}$ is minimum on the stable and unstable manifolds $W^{s/u}(\Gamma)$ of the hyperbolic periodic orbit $\Gamma$. Therefore $M_{S_0}$ attain a global minimum on the periodic orbit $\Gamma$. These results can be extended to the local neighbourhood of index one saddles of nonlinear systems due to the Moser's theorem \cite{moser1958generalization}. 

\begin{figure}[htbp]
	\begin{center}
		\includegraphics[scale=0.5]{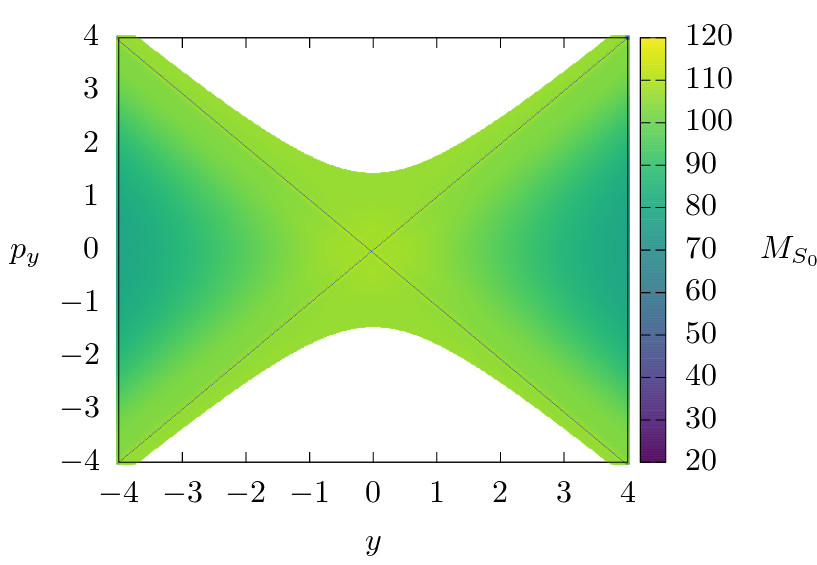} 
    \end{center}
    \caption{Lagrangian descriptor $M_{S_0}$ evaluated on the set of initial conditions on plane $y$--$p_y$ and $x=0$ for $\tau+=\tau_-=\tau=10$.  The region on white colour is a the forbidden region for the trajectories with $E=1$. The lines with the minimum value are $W^s(\Gamma)$ ($p_y=y$) and $W^u(\Gamma)$ ($p_y=-y$). To avoid large values of kinetic energy $T$ and the action $S_0$, the calculation of the trajectories stop when integration time is completed or when the particle rich the the circumference in the configuration space with radius $r=10$ with centre at the origin.
    }
    \label{fig:ld_saddle}
\end{figure}

\begin{figure}[htbp]
	\begin{center}
		\includegraphics[scale=0.5]{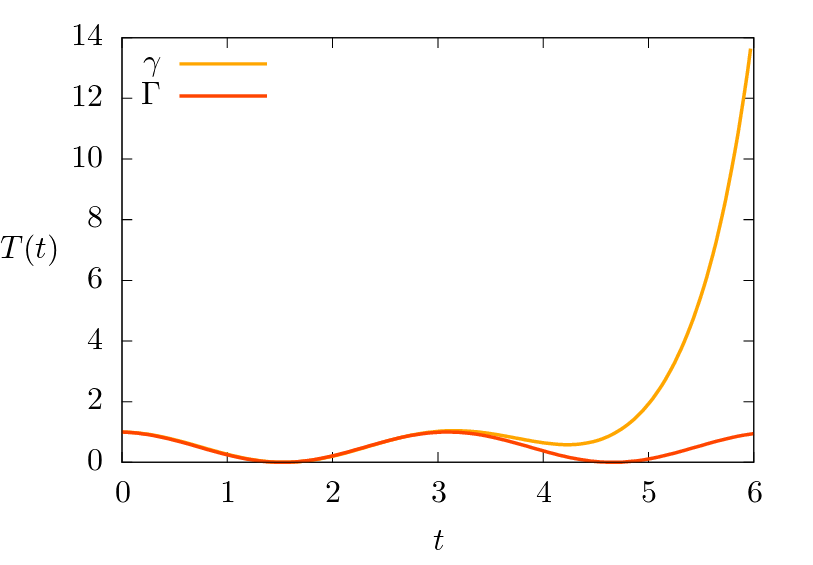}
    \end{center}
    \caption{ Kinetic energy $T(t)$ as a function of time for the the unstable hyperbolic periodic orbit $\Gamma$ and a trajectory $\gamma$ with slightly different initial conditions.  The kinetic energy for $\Gamma$ is a periodic function meanwhile the kinetic energy for $\gamma$ grows unbounded. The values of the action $S_0$ for each trajectory is twice the area under their corresponding $T(t)$ curve.
    }
    \label{fig:kinetic_trajectories_saddle}
\end{figure}

Intuitively, it is easy to understand this result considering the shape of the potential energy surface $V(x,y)$ around the index one saddle point and the trajectories around the stable manifold $W^s (\Gamma)$. For energies $E>0$, the unstable hyperbolic periodic orbit $\Gamma$ oscillates, and its kinetic energy $T$ is a periodic function of the time. Almost all the trajectories in a neighbourhood of the unstable hyperbolic periodic orbit $\Gamma$ separate from it and its kinetic energy grows due to the shape of $V(x,y)$ in the $y$-direction and the conservation of $E$, see figures \ref{fig:trajectories_saddle} and \ref{fig:kinetic_trajectories_saddle}. However, only the trajectories in $W^s (\Gamma)$ converge to $\Gamma$ and remain bounded. Then, their kinetic energy $T$ converge to a periodic function and the integral respect to the time of $2T$, the action $S_0$, is minimum for the trajectories on $W^s(\Gamma)$. As a result, the Lagrangian descriptor has a minimum in the stable and unstable invariant manifolds, and the global minimum reveals the position of their intersection in $\Gamma$, see figure \ref{fig:ld_saddle}. 

It is possible to generalise this result for multidimensional systems with index one saddle point in the potential energy using analogous arguments. In this case, the system has a Normally Hyperbolic Invariant Manifold (NHIM) associated with the saddle point in the multidimensional potential energy \cite{wig2016,wiggins3} . The Lagrangian descriptor based on the action is minimum in the invariant stable and unstable manifolds of the NHIM, and the global minimum in the intersection between the stable and unstable manifolds, give us the position of the NHIM in the phase space \cite{Wiggins2019finding}. The proof of this result is direct from the previous considerations; we need to consider more oscillatory degrees of freedom in the argumentation.

\begin{figure}[htbp]
	\begin{center}
		\includegraphics[scale=0.5]{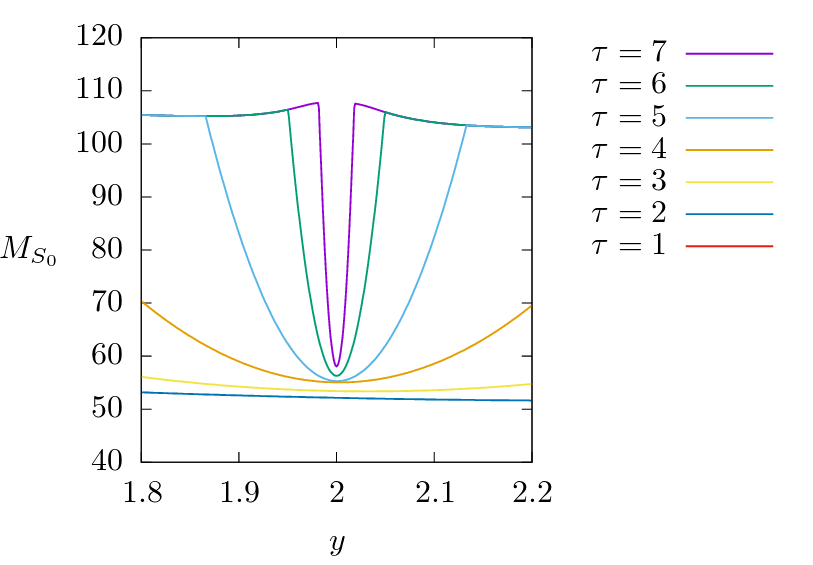}
    \end{center}
    \caption{Lagrangian descriptor $M_{S_0}$ evaluated on the set of initial conditions on a line $p_y=2$ and $x=0$ for different times $\tau$. The minimum value are intersection of the $W^u(\Gamma)$ with the set of initial conditions. To avoid large values of kinetic energy $T$ and the action $S_0$, the calculation of the trajectories stop when integration time is $\tau$ or the particle rich the the circumference in the configuration space with radius $r=10$ with centre at the origin. For $\tau>4$ the extremes of the curves does not change because their corresponding trajectories rich the circumference.}
    \label{fig:ld_linea_saddle}
\end{figure}

\newpage
\section{ Phase space analysis of a chaotic Hamiltonian using $M_{S_0}$}
\label{sec:Numerical example}

In this section, we illustrate the Lagrangian descriptor's capabilities $M_{S_0}$ to obtain the phase space structure of a chaotic system. We compare the Lagrangian descriptor plots with the corresponding Poincare maps with the same initial conditions. Let us consider a chaotic 2-DoF Hamiltonian proposed to qualitatively study a special kind of chemical reaction dynamics \cite{Carpenter2014}. The model has been analysed in detail using the Poincare map in \cite{katsanikas2018,katsanikas2019}.

\begin{figure}[htbp]
	\begin{center}
		\includegraphics[scale=0.5]{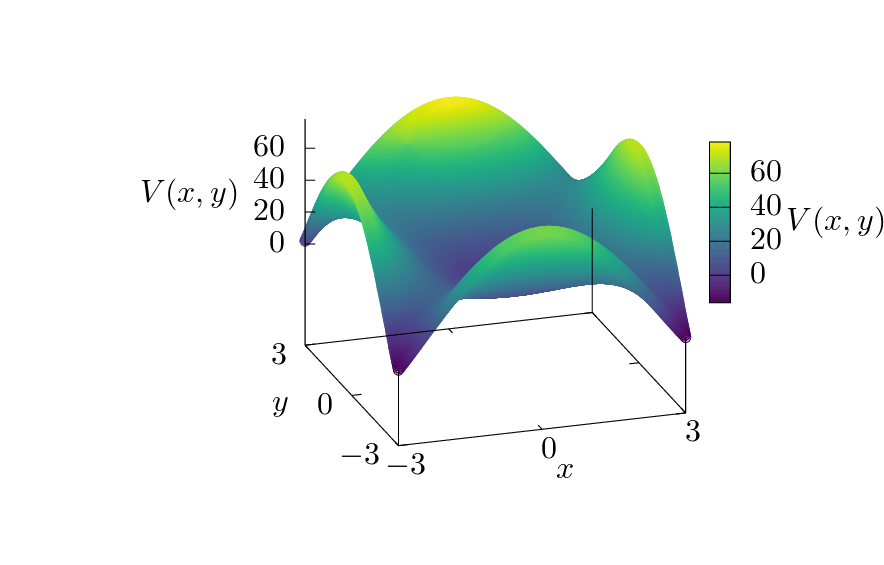}
    \end{center}
    \caption{Caldera potential energy $V(x,y)$ surface. The surface has a minimum has a central minimum and four index one saddle points at the corners. }
    \label{fig:V}
\end{figure}

\begin{figure}[htbp]
	\begin{center}
		\includegraphics[scale=0.45]{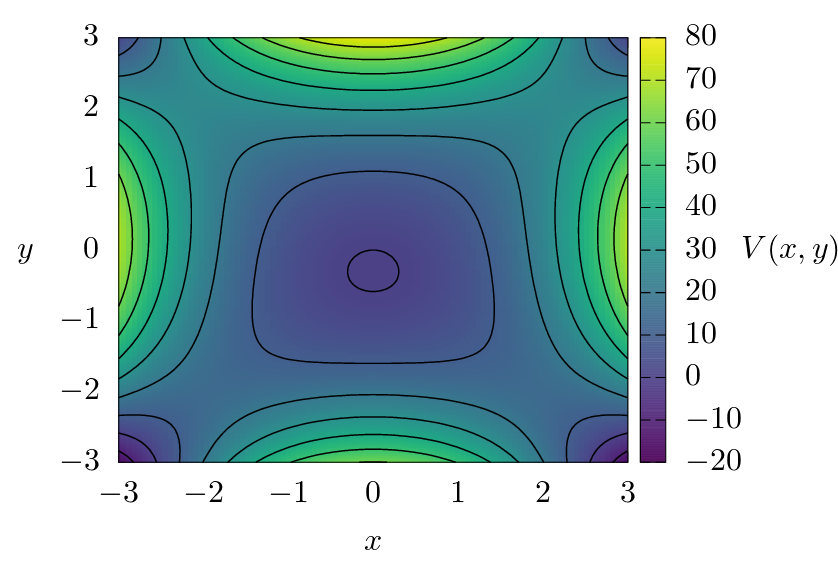}
    \end{center}
    \caption{Potential energy $V(x,y)$ in colour scale. The black lines are equipotentials.  }
    \label{fig:Vc}
\end{figure}

The Hamiltonian function of the system is 

\begin{equation}
H(x,y,p_x,p_y) = T(p_x,p_y) + V(x,y)  =   \frac{p^2_x}{2m} + \frac{p^2_y}{2m} + c_1( x^2 + y^2 ) + c_2 y - c_3 ( x^4 + y^4 - 6 x^2 y^2 ) 
\label{eq:V}
\end{equation}

\noindent where the mass of the particle is $m=1$ and the potential parameters are $c_1=5$, $c_2=3$, and $c_3=-3/10$. The potential energy $V(x,y)$ is symmetric with respect to the $y$-axis. This potential has a central minimum and four index one saddles around it, see figures \ref{fig:Vc} and  \ref{fig:V}. The shape of this potential energy surface is similar to a volcano's caldera. More information about this potential and the chemical reaction dynamics is in \cite{Carpenter2014} and references therein. 

\begin{figure}[htbp]
	\begin{center}
		A)\includegraphics[scale=0.37]{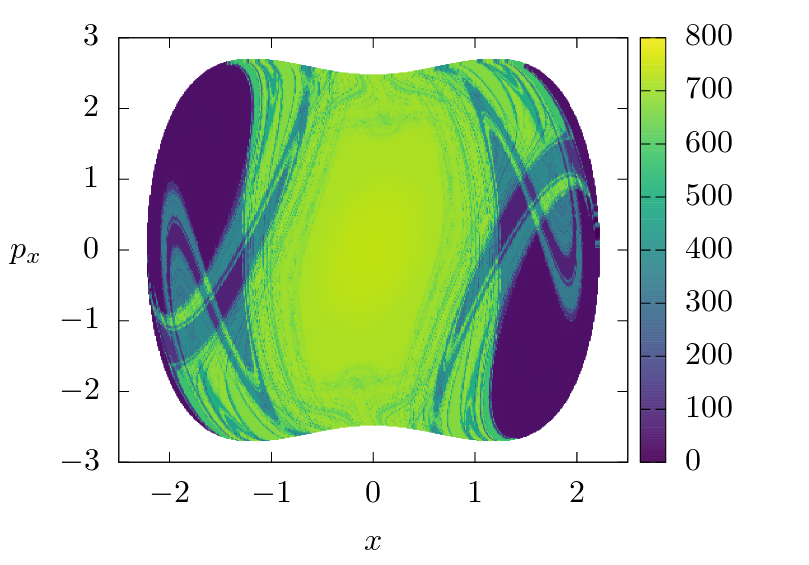}
		B)\includegraphics[scale=0.35]{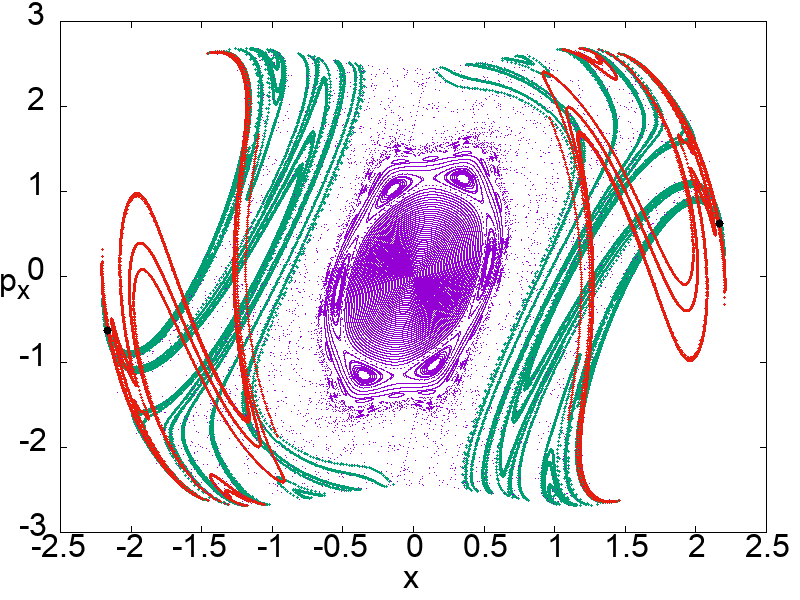}
		C)\includegraphics[scale=0.37]{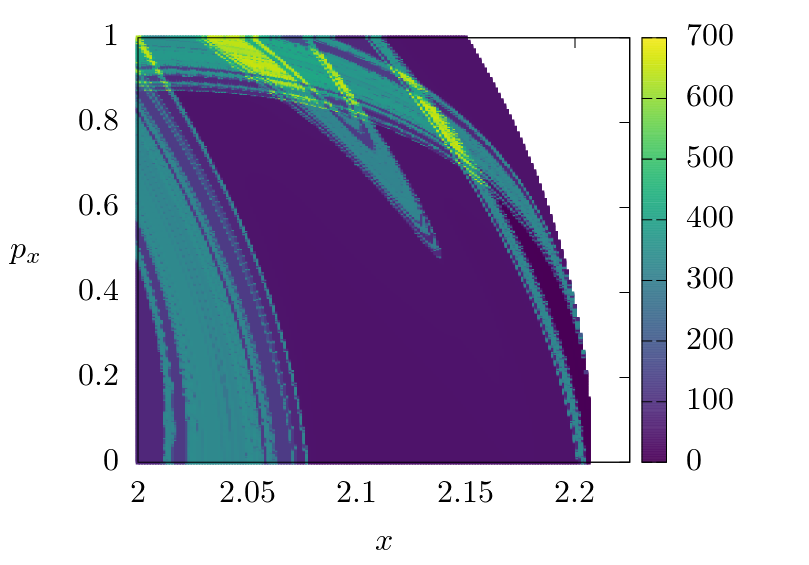}
		D)\includegraphics[scale=0.35]{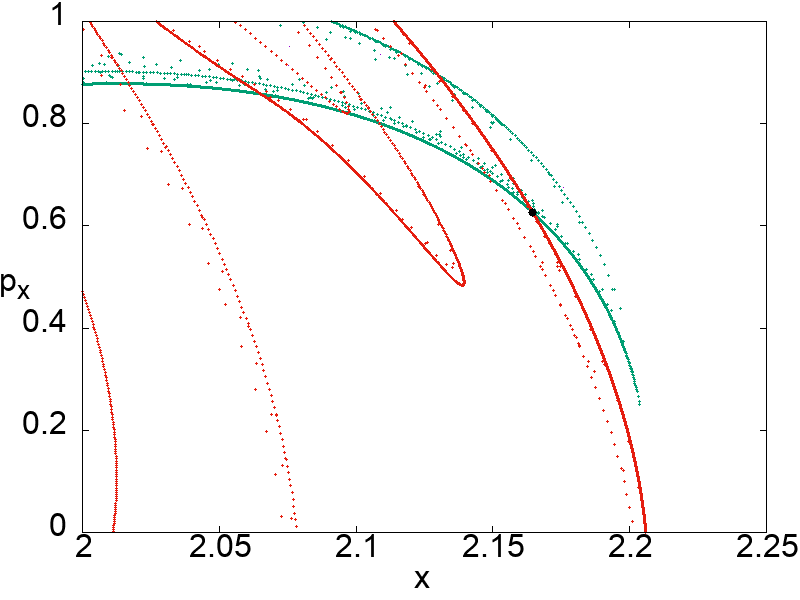}

	\end{center}
	\caption{ Lagrangian descriptor $M_{S_0}$ and Poincare map with mixed phase space for $E=17$. The Poincare section is the canonical conjugate plane $x$--$px$ with $y = -1.8019693$ and $p_y>0$. The trajectories to calculate $M_{S_0}$ stop when the integration time is $\tau = 20$ or the particles reach the a circumference in the configuration space with radios $r=3$ and centre at the origin.}
	\label{fig:phase_space_E17} 
\end{figure}

 Let us consider two values for the energy $E$ such that the phase space of the system is unbounded and some KAM islands are present. For the first value of the energy, $E=17$, we take the canonical conjugate plane $x$--$p_{x}$ with $y= -1.8019693$. The component of the momentum $p_y>0$ is determined by the conservation of the energy. The figure \ref{fig:phase_space_E17} shows Lagrangian descriptor plots on the left side and their corresponding Poincare maps on the right side. The similarities between the patterns on both kinds of plots are clear.

\begin{figure}[htbp]
	\begin{center}
		A)\includegraphics[scale=0.5]{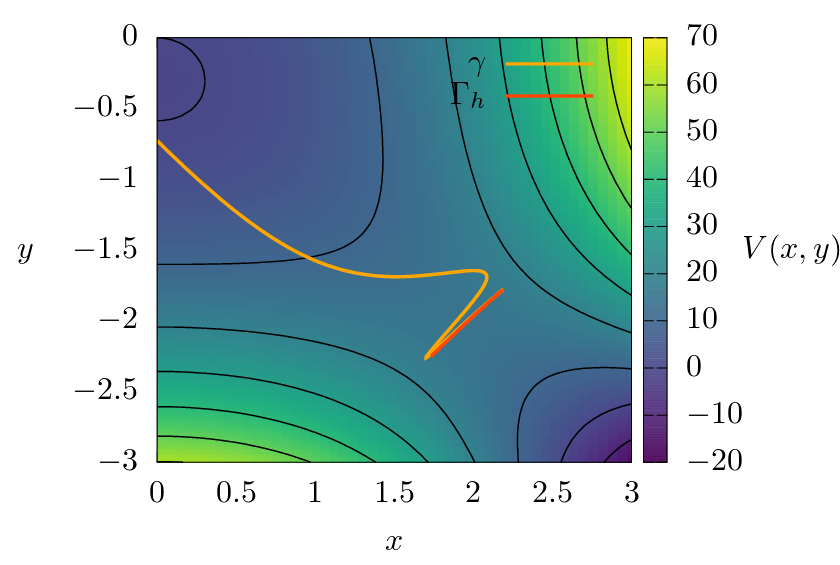}
	\end{center}
	\caption{ Projection in configuration space of the periodic orbit $\Gamma_h$ associated to a saddle point in the potential energy surface and a trajectory $\gamma$ with slightly different initial conditions. 
	The potential energy $V(x,y)$ is on the background. The periodic orbit $\Gamma_h$ intersects the Poincare surface $x$--$p_x$ with $y = -1.8019693$ on the figure \ref{fig:phase_space_E17} D) at the black point $(x,p_x)=(2.16457,0.626450)$.  }
	\label{fig:trajectories_E17} 
\end{figure}

The Poincare map in panel B) of figure \ref{fig:phase_space_E17} shows the intersections of some trajectories with the Poincare plane. In this panel, we can see a central KAM island with violet points, the chaotic region between the closed invariant curves, the transient chaotic region outside the KAM islands where the dynamics is complicated just for some time, and the unbounded region that is typical in open Hamiltonian systems, some examples of transient chaotic systems  with 2 and 3 DoF are in  \cite{Tel_book,Tel2015,Janosi2019,Gonzalez2012,Gonzalez2014}. The curves on green and red are the intersections of the unstable manifold $W^s(\Gamma_h$) and stable manifold $W^u(\Gamma_h$) of the hyperbolic periodic orbit $\Gamma_h$ and its symmetric periodic orbit (with respect to the $y$-axis) with the Poincare plane. This periodic orbit and its symmetric periodic orbit are Lyapunov orbits associated with the low right and low left index one saddle points in the potential energy surface. The projection of orbit $\Gamma_h$ in the configuration space is in figure \ref{fig:trajectories_E17}. In the panel D) of figure \ref{fig:phase_space_E17} there is a magnification of the region where $\Gamma_h$ intersects with the Poincare plane at the black point around $(x,p_x)=(2.16457,0.626450)$. 

The Lagrangian descriptor plots show the regions of the phase space where the trajectories are unbounded (dark-blue), transient chaotic (green), and close to regular (bright-yellow). There is a good agreement between the Poincare maps and the Lagrangian descriptor plots on the KAM islands and transient chaotic areas generated by the tangle of the hyperbolic periodic orbit $\Gamma_h$ and its symmetric periodic orbit.

Let us consider the origin of the signature of phase space objects in the Lagrangian descriptor plots using the conservation of the energy and the behaviour of the trajectories. The trapped regions inside the KAM islands have large values of $M_{S_0}$. The trapped trajectories in the KAM island remain all the time in a region nearby the global minimum of the potential energy $V$, see figures \ref{fig:V} and \ref{fig:Vc}. That means that for large values of time $t$ the kinetic energy $T(t)$ of trapped trajectories is larger than the kinetic energy of the trajectories that spend only a lapse close to the minimum of $V$ and then reach an exit of the caldera. Thus, the Lagrangian descriptor's values $M_{S_0}$ for the trapped trajectories in the island are larger than the values for the trajectories that reach an exit. 

\begin{figure}[htbp]
	\begin{center}
		A)\includegraphics[scale=0.5]{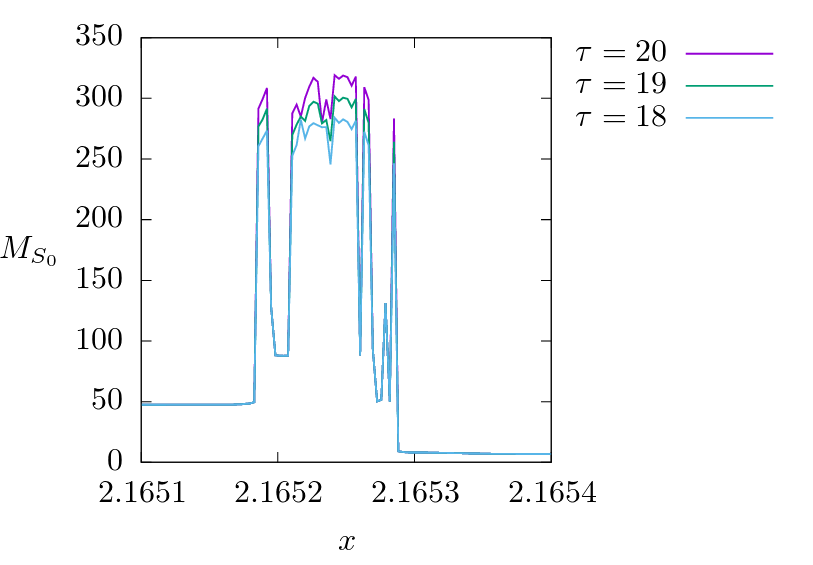}
	\end{center}
	\caption{ Lagrangian descriptor $M_{S_0}$ for different times $\tau = 18,19,20$. The set of initial conditions is a line segment on the domain of the figure \ref{fig:phase_space_E17} C). The first abrupt jump on the right side corresponds to the intersection of the stable manifold $W^s(\Gamma)$ with the set of initial conditions. All the trajectories with initial conditions on the right reach the circumference with $r=3$. For this reason we only see an abrupt jump on $M_{S_0}$.  }
	\label{fig:ld_line_E17} 
\end{figure}

\begin{figure}[htbp]
	\begin{center}
		A)\includegraphics[scale=0.37]{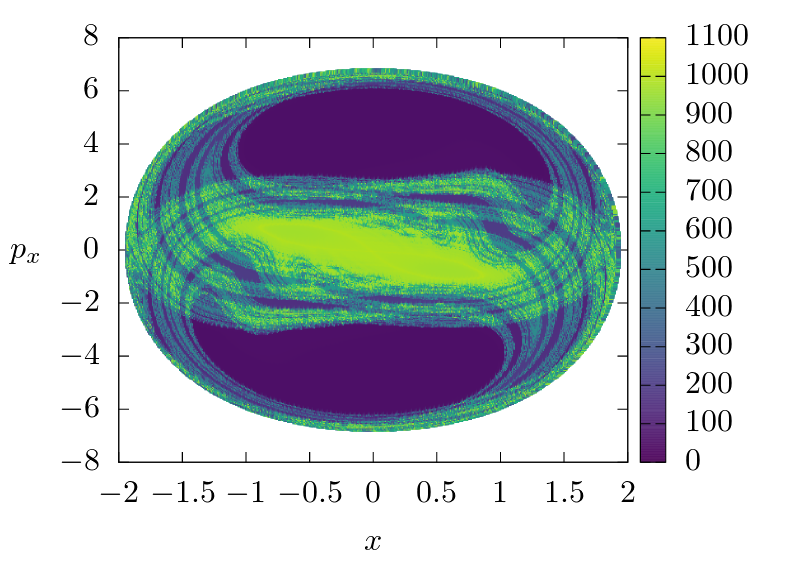}
		B)\includegraphics[scale=0.35]{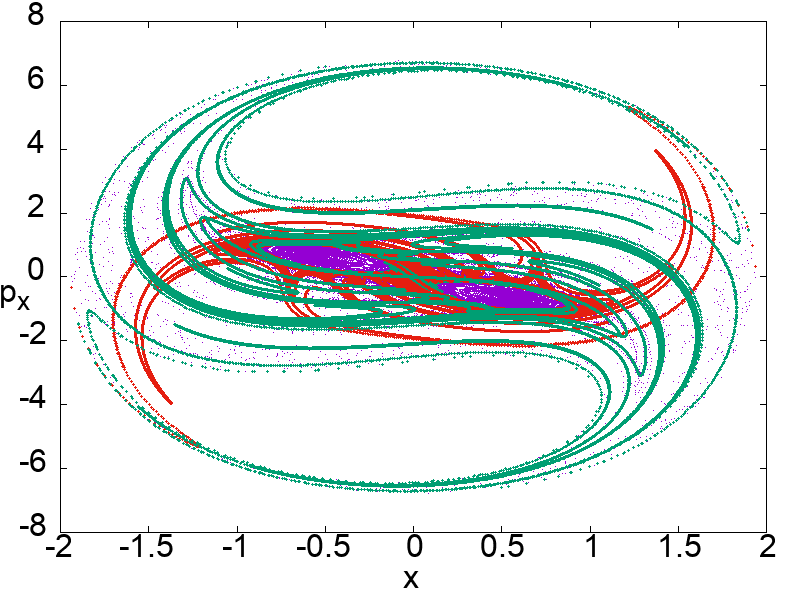}
		C)\includegraphics[scale=0.37]{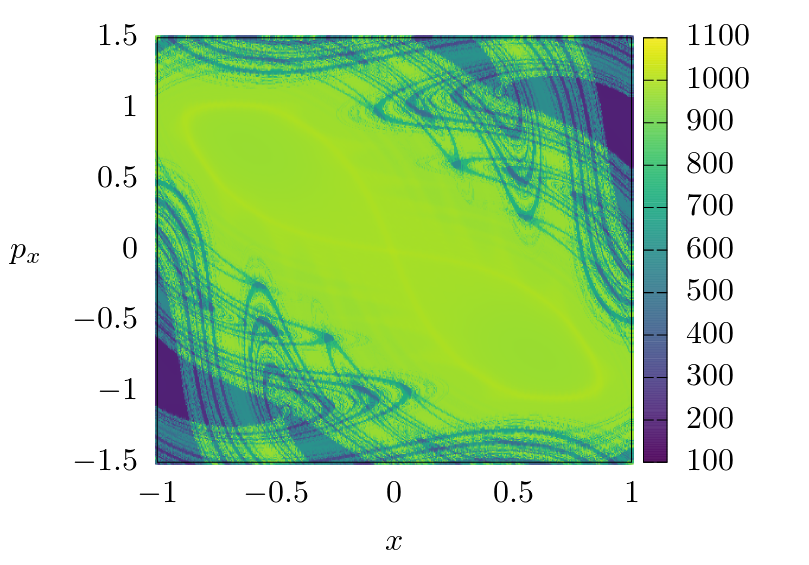}
		D)\includegraphics[scale=0.35]{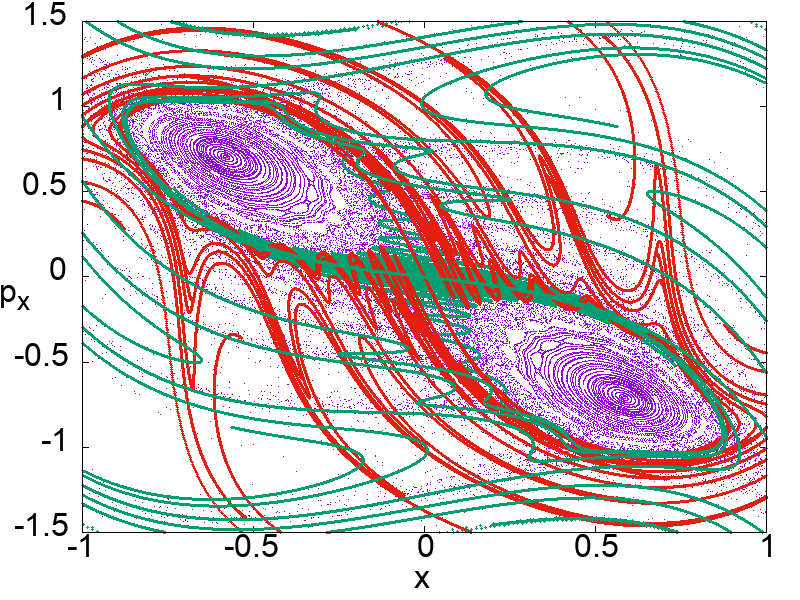}

	\end{center}
	\caption{ Lagrangian descriptor $M_{S_0}$ and Poincare map with mixed phase space for $E=23$. The Poincare surface of section is the canonical conjugate plane $x$--$p_x$ with $y = 0$ and $p_y>0$. The maximal integration time for the Lagrangian descriptor plots is $\tau = 20$. }
	\label{fig:phase_space_E23} 
\end{figure}

\begin{figure}[htbp]
	\begin{center}
    \includegraphics[scale=0.5]{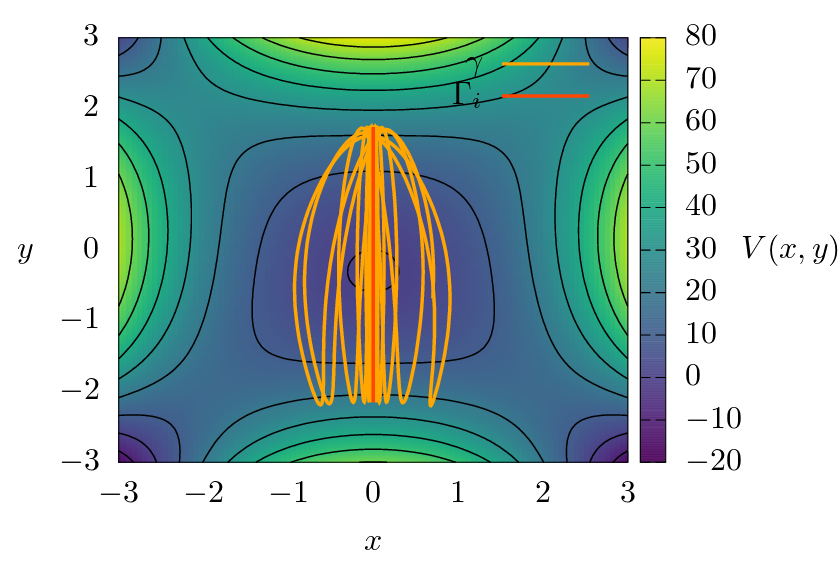}
	\end{center}
	\caption{ Projection on configuration space of the periodic orbit $\Gamma_i$ corresponding to the point in the centre of the figures in \ref{fig:phase_space_E23} and a trajectory $\gamma$ with slightly different initial conditions. The potential energy $V(x,y)$ is the background in colour scale with some equipotential lines on black. The periodic orbit $\Gamma_i$ oscillates on $y$-direction between two equipotential lines.
	}
	\label{fig:trajectories_E23} 
\end{figure}

\begin{figure}[htbp]
	\begin{center}
		A)\includegraphics[scale=0.5]{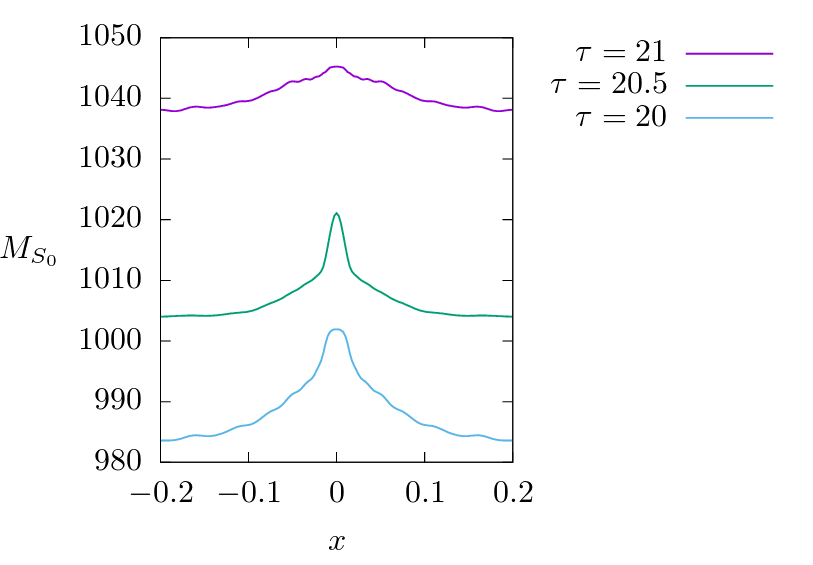}
	\end{center}
	\caption{ Lagrangian descriptor $M_{S_0}$ for different times $t$. The set of initial conditions is a line segment on the figure \ref{fig:phase_space_E23} D) with $p_x=0$. The maximum corresponds to the intersection of the inverse hyperbolic periodic orbit $\Gamma_i$ with the set of initial conditions.}
	\label{fig:ld_line_E23} 
\end{figure}

The stable and unstable manifolds of the hyperbolic periodic orbit $\Gamma_h$ generate abrupt changes in the Lagrangian descriptor's values. To see this property, let us consider a line of initial conditions that intersect $W^s(\Gamma_h)$. The line is contained on the domain of panel C) of figure \ref{fig:phase_space_E17} and has $p_x=0.62$. The results for different integration times are in figure \ref{fig:ld_line_E17}. We can see how the jumps are generated when $\tau$ is increased. The first jump on the right side corresponds to the intersection of the stable manifold $W(\Gamma_h)^s$ with the set of initial conditions. We see a jump and not a soft minimum like in the example in the previous section due to the large integration time and the stop condition for trajectories that reach the circumference outside the caldera. The trajectories on the right side of the $W(\Gamma_h)^s$ reach the circumference and calculation of their $M_{S_0}$ stops.


As a second example, let us consider another value of the energy where the phase space has a different structure. The figures in panels B) and D) of figure \ref{fig:phase_space_E23} show the Poincare map corresponding to the canonical conjugate plane $x$--$p_x$ with $y=0$, $p_y>0$ and $E=23$. The central black point corresponds to the intersection between the Poincare plane and an inverse hyperbolic periodic orbit $\Gamma_i$. Figure \ref{fig:trajectories_E23} shows the projection of $\Gamma_i$ in the configuration space.  This periodic orbit belongs to the family of periodic orbits of the central minimum \cite{katsanikas2018}.  The red and green lines on the Poincare maps of figure \ref{fig:phase_space_E23} are the stable and unstable manifolds $W^s(\Gamma_i)$ and $W^u(\Gamma_i)$.

In this case, the results are different from the results of the integrable example where the Lagrangian descriptor has a minimum at hyperbolic periodic orbit $\Gamma$. For the inverse hyperbolic periodic orbit $\Gamma_i$, the Lagrangian descriptor $M_{S_0}$ have a maximum at the intersection between the inverse hyperbolic periodic orbit and the set of initial conditions. Figure \ref{fig:ld_line_E23} shows $M_{S_0}$ evaluated on a line of initial conditions that intersect the $\Gamma_i$ for different values of the integration time $\tau$. To explain this different behaviour, with respect to the behaviour of $M_{S_0}$ for the hyperbolic periodic orbit $\Gamma$ associated with the saddle in the potential energy surface, we consider the conservation of the energy and the shape of the potential energy surface around the projection of $\Gamma_i$ again. The projection of $\Gamma_i$ in the configuration space is at the minimum of $V(x,y)$ with respect to the $x$-direction. The trajectories in the neighbourhood of $\Gamma_i$ have larger values of $V(x,y)$ and their kinetic energy $T$ is smaller than the kinetic energy for $\Gamma_i$. Consequently, the Lagrangian descriptor has a maximum on $\Gamma_i$ and its stable and unstable manifolds $W^s(\Gamma_i)$ and $W^u(\Gamma_i)$.

\newpage

\section{Conclusions and remarks}
\label{sec:Conclusions and remarks}

We construct a natural phase space structure indicator for Hamiltonian systems based on the Maupertuis' action $S_0$. For this construction, it is necessary that the kinetic energy is a quadratic function of the generalised velocities, and its potential energy is only a function of the generalised coordinates. The simplest way to calculate the Lagrangian descriptor $M_{S_0}$ is with the integral of the kinetic energy with respect to the time of the trajectories of the system. 
The Lagrangian descriptor based on the action is a convenient tool to study the phase space of open Hamiltonian systems. The energy conservation is essential to interpret the values of $M_{S_0}$ and link them with the potential energy $V$. There is a simple relationship between its values and the different phase space regions. The trapped regions correspond to large values of $M_{S_0}$ and the unbounded regions to smaller values of $M_{S_0}$. 

With the action as a phase space structure indicator, it is possible to identify regular regions, unbounded regions, resonant islands and the transient chaotic sea around them formed by homoclinic and heteroclinic tangles. In the regions where the dynamics is confined to a finite region on the phase space, the values of $M_{S_0}$ change smoothly for large integration times and is clear how to identify the central periodic orbits in the islands. However, we can always find stable periodic orbits on the KAM islands centres with the Poincare map. Therefore, the Lagrangian descriptor and the Poincare maps are complementary tools to reveal the phase space structure.

We find that $M_{S_0}$ has a minimum value on their stable and unstable manifolds of the hyperbolic periodic orbits. On the other hand, this Lagrangian descriptor has a maximal value for the stable and unstable manifolds for the inverse hyperbolic periodic orbits. These results are intuitive if we consider the conservation of the energy and the trajectories' behaviour in the periodic orbits' neighbourhood. It is possible to generalise these results for NHIMs and their stable and unstable manifolds in systems with more dimensions.

In some situations with unbounded negative potentials, it is convenient to stop calculation of the Lagrangian descriptor's trajectories. In this way, the Lagrangian descriptor reveals only the phase space structure in one particular region. It is important to consider the stop of the trajectories to interpret the Lagrangian descriptor. For the Lagrangian descriptor plot on figures \ref{fig:phase_space_E23} and \ref{fig:ld_line_E23}, the stop of the integration of the trajectories generates an abrupt jump that reveals $\Gamma_h$.

\section{Acknowledgments}
\label{sec:Acknowledgments}

We acknowledge Stephen Wiggins for discussions about the action and the support of EPSRC Grant no. EP/P021123/1.

\bibliography{bibliography}

\end{document}